\documentclass[aps, prd, preprint,12pt]{revtex4-1}
\usepackage{amsfonts}
\usepackage{amssymb}
\usepackage{amsmath}
\usepackage{graphicx}

\begin{document}
\title{Positive-Energy  Dirac Particles and Dark Matter}
\author{Eugene Bogomolny}
\affiliation{Universit\'e Paris-Saclay, CNRS, LPTMS, 
 91405 Orsay, France}
\date{\today}

\begin{abstract}
The  relativistic positive-energy wave equation proposed by P. Dirac in  1971 
 is an old but largely forgotten subject. The purpose of  this note is to speculate  that  particles described by this equation (called here  Dirac particles) are  natural candidates for the dark matter. The reasoning is based on a fact that  the internal structure of such particles simply prohibits  their  interaction with electromagnetic fields (at least with the minimal coupling) which is exactly what is required  for dark matter.  Dirac particles have quite unusual properties. In particular, they  are transformed by an infinite-dimensional representation of the homogeneous Lorentz group, which clearly distinguishes them  from all known elementary particles described by  finite-dimensional representations and  hints to a physics beyond  the Standard Model.   To  clarify the topic, a brief  review of the main features of {the above-mentioned}  Dirac equation is given.  
\end{abstract}

\maketitle

 \section{Introduction}  A tremendous number of various astrophysical observations collected during the last century are explained by postulating the existence of invisible (i.e., not interacting with electromagnetic fields) matter.  Such a hypothetical substance goes under the name of  dark matter, and according to different estimates, it constitutes  $\sim85\%$ of  the total  matter content of the universe (see, e.g.,  reviews~\cite{supersymmetricDM,particleDM,nature} and references therein).  Though dark matter is a predominant ingredient of the universe, its composition remains unknown. A great  number of different candidates have been  proposed  \cite{supersymmetricDM,particleDM,nature}, but none have of course been confirmed so far.   

The purpose of this short  note  is to suggest  a (seemingly unexplored)  possibility  that dark matter is made up of particles  obeying the  so-called new Dirac equation proposed by P. Dirac  \cite{Dirac_I} 40 years after his famous (old) equation \cite{Dirac}.  
Both equations are relativistic wave equations. The usual (old) Dirac equation  describes particles with positive energy  as well as antiparticles with negative energy and is one of the best known equations in physics. The principal feature of the new equation is that   all its solutions  have only  positive energies (thus, antiparticles are absent or, at least,  not described by the same equation). 
 For relativistic equations, such property is unusual. The relativistic invariance implies {that}
  $E^2=\vec{p}^{\,2}+m^2$, {and in general, there exist solutions with positive as well as negative energies: $E=\pm\sqrt{ \vec{p}^{\,2}+m^2}$.  Though today negative energy solutions simply indicate the existence of antiparticles, the search for relativistic equations with only positive energies continues to  persist (see, e.g., \cite{cirilo} and references therein). }

It was  E. Majorana \cite{majorana, fradkin} who  first  constructed a relativistic wave equation with only positive energy solutions. It appears that it becomes {feasible} provided one considers particles transformed by an infinite-dimensional representation of the homogeneous  {(i.e., without translations)} Lorenz group. Usual relativistic particles (scalars, spinors, vectors, etc.) realise  finite-dimensional representations of the Lorentz group,  and their equations necessarily have positive and negative energy solutions.  {The positive-energy} Majorana equation formally has   form of the usual Dirac equation: 
\begin{equation}
\big (\mathbf{\Gamma}^{\mu}\partial_{\mu}- m  \big )\mathbf{\Psi}(x)=0
\label{old_majorana}
\end{equation} 
but with the difference that its eigenfunctions $\mathbf{\Psi}(x)$ have {an} infinite number of components (labeled by two non-negative integers)  and matrix $\mathbf{\Gamma}^{\mu}$ is an infinite-dimensional matrix  derived from the requirement of the relativistic invariance (whose explicit form could be obtained from expressions in Section~\ref{additional_equations}).  This  equation explicitly describes  an infinite number of particles  with all integer and half-integer spins and masses decreasing with spin: $m(j)=m/(j+1/2)$ \cite{majorana, fradkin}.  Equation \eqref{old_majorana} should not be confused with another much better-known Majorana equation   for self-conjugate particles with spin-$\tfrac{1}{2}$ \cite{usual_majorana}.  

{ The distinction  between finite-dimensional  and  infinite-dimensional  representations of a group is that for the former, the  action  of group generators mixes  a finite number of wave components (e.g., four for the usual spinors) but for the latter, such action requires an infinite number of components. Mathematically, it is well established  that the homogenous Lorentz group has both finite and infinite-dimensional representations (see, \mbox{e.g., \cite{harish, gelfand}} and references therein).  All known elementary particles are classified according  to finite-dimensional representations of this group,  yet   particles associated  with its infinite-dimensional  representations  are not usually taken into account.}

In the new equation, Dirac found an elegant way (see below) to select one state from an infinite tower of Majorana solutions.  Particles described by the new Dirac equation  are also transformed by an infinite-dimensional representation of the Lorenz group, which markedly  differentiates them  from all known elementary particles. {Such particles can be called positive-energy Dirac particles, but for the sake of brevity, in this note, they are referred to as Dirac particles.} 

Another characteristic  feature of these Dirac particles is a nontrivial  fact that they cannot interact with electromagnetic fields (at least with the minimal coupling). Such a no-interaction theorem is related with the internal   structure of the new Dirac equation and is not imposed by simply postulating that the charge of Dirac particles equals zero. This propriety emerges  as a byproduct of Dirac's construction, and in the past has been  considered as a serious defect of the new Dirac equation, but it is exactly what makes Dirac particles a natural candidate for dark matter, which by definition is a substance not interacting  with electromagnetic fields. 

{In the literature, there are articles that relate Dirac particles and dark matter (see, e.g., \cite{dirac_dark_matter}, besides others), but in these papers, the term 'Dirac particles' indicates solely spin-$\tfrac{1}{2}$ particles obeying the usual Dirac equation \cite{Dirac}. Dirac particles discussed here are of a different nature as they satisfy   the new  Dirac equation \cite{Dirac_I} and are transformed by an infinite-dimensional representation of the Lorentz group.  To the author's knowledge, such particles were not considered previously as candidates for dark matter.}

The purpose of the paper is to attract wider attention to  this potentially important  subject and to serve as a brief introduction to Dirac particle properties.  As such  issues  are largely undeveloped, the discussion below is restricted only to principal  features of Dirac particles leaving aside many important questions.  
 
 The plan of the paper is the following. In Section~\ref{new_Dirac_equation},  the main properties of the new Dirac equation are discussed. Section~\ref{infinite_dimensional_representation} is devoted to the investigation of the relativistic invariance of this equation and to the meaning of {the} infinite-dimensional representation of its solutions.  By construction, the new Dirac equation is an overdetermined system of equations, and consequently,  one can derive additional equations that are indicated in Section~\ref{additional_equations}. Section~\ref{no_interaction_theorem} deals with  the no-electromagnetic-interaction property  of the new Dirac equation, which is of primary  importance for the conjectured relation of  Dirac particles with dark matter. In Section~\ref{gravity_Dirac_particles}, the  interaction of Dirac particles with gravitational fields is shortly  treated. Section~\ref{conclusion} gives an overview of the discussed  topics. For clarity, certain useful formulas are 
 {presented}  in Appendix~\ref{app_A}, and for completeness, the tetrad formalism is recalled  in Appendix~\ref{app_tetrad}.


 \section{New Dirac Equation}\label{new_Dirac_equation} 
 
 The usual Dirac equation \cite{Dirac} describes  relativistic fermions with spin $\tfrac{1}{2}$  and mass $m$ and is one of the most important and the best-known equations of mathematical physics.   Much less  attention has been attracted  {to}  another Dirac work \cite{Dirac_I}, written  about 40 years after the famous paper \cite{Dirac}, in which  a different relativistic equation (called the new Dirac equation) was proposed and analysed. 
 
The new equation  has the  functional form similar to  the usual Dirac equation:
\begin{equation}
\left ( \gamma^{\mu}\partial_{\mu}-m \right )\Psi =0
\label{dirac}
\end{equation}
but with an important difference in that  column-vector $\Psi$  depends on two sets of variables:   four  space-time coordinates $x^{\mu}$ with $\mu=0,1,2,3$ and two 'internal'  (auxiliary) variables $q_1$  and $q_2$, arranged in the following manner:
\begin{equation}
\Psi(x,q)=Q \, \Phi(x, q)\, ,\qquad Q= \left ( \begin{array}{c} q_1\\q_2\\ q_3\\q_4\end{array}\right )
\label{new_dirac}
\end{equation}
where components $q_3$ and $q_4$ are the momenta  in the 'internal' space
\begin{equation}
 q_3=-i\frac{\partial}{\partial q_1}\, ,\qquad  q_4=-i\frac{\partial}{\partial q_2}\, .
\end{equation}

The choice of  such column-vector $Q$   corresponds to  the  following commutation relations: 
\begin{equation}
[q_{j},q_k]=i\beta_{j k}\, ,\qquad \beta=\left ( \begin{array}{cc} {0}& {1}\\- {1}& {0} \end{array}\right ). 
\label{commutators}
\end{equation}

As for the usual Dirac equation,  $\gamma^{\mu}$ are $4\times 4$ matrices obeying  
$\gamma^{\mu}\gamma^{\nu}+\gamma^{\nu}\gamma^{\mu}=2\eta^{\mu \nu}$ 
with the metric tensor of the Minkowski space $\eta^{\mu \nu}=\mathrm{diag}(-1,1,1,1)$ (cf. \eqref{usual_gamma}). 
 
Function  $\Phi(x, q)$ in \eqref{new_dirac} depends on  both sets of coordinates (i.e. on six variables), and it is the only one  unknown function.  Thus, the new Dirac equation corresponds to an overdetermined system of four equations {in} one unknown function:
\begin{equation}
\hat{P}_j \, \Phi(x,q)=0\, , \qquad \hat{P}_j =(\gamma^{\mu} \partial_{\mu}-m)_{j k} q_k\, .
\label{P_ops}
\end{equation}

To ensure the consistency of these equations,  it is necessary that the commutator of any two $\hat{P}_j $ equals zero. Using \eqref{commutators} one finds
\begin{equation}
[\hat{P}_j ,\,  \hat{P}_k]= \big ( (\gamma^{\mu} \partial_{\mu}-m)i \beta(\gamma^{\mu \,T} \partial_{\mu}-m)\big )_{j k}
\end{equation}
where $\gamma^{\mu\, T}$ are {the} transposed gamma-matrices. In the new Dirac equation,  these matrices have to be  chosen to fulfil the requirement
\begin{equation}
\beta\,\gamma^{\mu \,T}=-\gamma^{\mu}\,\beta\, .
\label{condition}
\end{equation}

By construction, matrix $\beta$ is an antisymmetric matrix (cf. (\ref{commutators})),  and this requirement means that $\beta \gamma^{\mu}$ are symmetric matrices. 
 
Under these conditions,  the above commutator becomes 
\begin{equation}
[\hat{P}_j \,\hat{P}_k]=i \beta_{j k} \big (-\eta^{\mu \nu}\partial_{\mu}\partial_{\nu} +m^2\big )
\end{equation}
which implies that all {four equations} 
 (\ref{P_ops}) are consistent iff $\Phi(x,q)$ as a function of $x$ obeys the Klein--Gordon equation:
\begin{equation}
\big (-\eta^{\mu \nu}\partial_{\mu}\partial_{\nu} +m^2\big )\Phi(x,q)=0\, .
\label{klein_gordon} 
\end{equation}

It is this {additional relation}  that differs the new Dirac equation from the Majorana equation~\eqref{old_majorana} \cite{majorana} and permits us to select one state from  an infinite number of Majorana solutions. 

It is plain that \eqref{klein_gordon} is a necessary condition for the existence of a  solution of  the new Dirac equation. It could be obtained by, e.g., the multiplication of \eqref{dirac} with  \eqref{new_dirac} by $Q^T\beta \, \big (\gamma^{\nu} \partial_{\nu}+m\big ) $ and using {the fact} that $Q^T\beta \,Q =2i$ (see \eqref{QGQ}). The sufficiently  of this equation is more subtle.  Usually, one refers to  the  Frobebius theorem (see, e.g., \cite{foliations} and references therein), which (in the simplest setting)  states that in order for  a system of linear differential equations to have  a solution, 
 it is necessary and sufficient that their commutators (more general, Lie brackets) lie in the their span. The important ingredient of this theorem is the fact that commutators cancel second derivatives, which is not valid for the  considered problem. Therefore, the statement that \eqref{klein_gordon} is the sufficient condition for the existence of solutions of the new Dirac equation requires additional investigations. 

It has been proved in \cite{Dirac_I} that these equations  determine  massive relativistic particles ({the} Dirac particles) with only positive energy and zero spin in the rest frame. In the next publication \cite{Dirac_II},  Dirac showed that  when {the} noncommutativity  of different variables is ignored, Dirac particles could be considered as vibrating spherical shells similar to \textit{Zitterbewegung} of particles obeying the usual Dirac equation.  
 
 The  simplest solution of the new Dirac equation  is the plane wave of the following form \cite{Dirac_I}:
 \begin{equation}
 \Phi(x,q)=C\exp\left (-\frac{1}{2}a q_1^2-\frac{1}{2}b q_2^2+ c q_1q_2+ip_{\mu} x^{\mu}\right )
 \label{plane_wave}
 \end{equation}
 where $p_0=\sqrt{\mathbf{p}^2+m^2}>0$, $C$ is a normalisation constant,  and coefficients $a,b,c$  are 
 \begin{equation}
 a=\frac{m+ip_1}{p_0+p_3}\, , \quad b=\frac{m-ip_1}{p_0+p_3}\, ,\quad c=\frac{i  p_2}{p_0+p_3}\, .
  \end{equation}
  
The positivity of Dirac particles' energy 
 is attested by the fact that negative energy solutions with $p_0=-\sqrt{m^2+\mathbf{p}^2} $ lead to non-normalised functions and have to be excluded. 
    

\section{Infinite-Dimensional Representation}\label{infinite_dimensional_representation}

 The relativistic (co)variance of the new Dirac equation means that  {a}  linear Lorentz transformation $x^{\prime \mu}=\Lambda^{\mu}_{\nu} x^{\nu} $  ($\eta^{\mu \nu}=\eta^{\sigma \rho} \Lambda^{\mu}_{\sigma} \Lambda^{\nu}_{\rho}$) of a solution 
\begin{equation}
\Phi(\Lambda x, q)=U(\Lambda ) \Phi(x,q) 
\label{covariance} 
\end{equation}
is also a solution of the same equations. Here, $U(\Lambda)$ is an operator acting on $q$ variables.

The inspection of Equation~\eqref{new_dirac}  shows that it will be the case if the following two conditions are fulfilled: 
\begin{equation}
U^{-1} \,Q \, U=V\, Q\, , \qquad   V\, \gamma \, V^{-1} = \gamma \, \Lambda
\label{two_eqs}
\end{equation}
where matrix $V\equiv V_{jk}(\Lambda)$ acts on spinor indices.  Indeed, under such changes,   Equation~\eqref{new_dirac} transforms linearly; thus,
\begin{equation}
\hat{P}\,\Phi(\Lambda x, q)=  U\,V\, \hat{P}\, \Phi(x,q) =0\, .
\end{equation}

The explicit expressions of $V$ and $U$ can easily be obtained from the infinitesimal Lorentz transformations 
\begin{equation}
\Lambda^{\mu}_{\nu}=\delta^{\mu}_{\nu}+a^{\mu}_{\nu}=\delta^{\mu}_{\nu}+\frac{1}{2} a_{cd} \big (I^{cd} \big)^{\mu}_{\ \nu} 
\label{Lorentz_transform}
\end{equation}
where $4\times 4$ antisymmetric matrix $a_{cd}$ describes  six parameters of the Lorentz transformation
and  matrices  $I^{cd}$ are generators of Lorentz transformation
\begin{equation}
 \big (I^{cd} \big)^{\mu}_{\ \nu} =\eta^{c \mu} \delta^{d}_{\nu} -\eta^{d \mu} \delta^{c}_{\nu}\, .
\end{equation}

These generators  obey the commutator relations of the Lie algebra of the Lorenz group
\begin{equation}
\big [ I^{cd},I^{kr} \big ]= \eta^{d k} I^{c r} +\eta^{c r} I^{d k} - \eta^{d r} I^{c k}-\eta^{c k} I^{d r}\, .
\label{Lorentz_generators}
\end{equation}

It is plain  that  $V$ and $U$ should be of the following  form: 
\begin{equation}
V=\exp \Big (\frac{1}{2} a_{cd}\Sigma^{cd}\Big ), \qquad  U=\exp \Big (\frac{i }{2} a_{cd}S^{cd}\Big ) .
\end{equation}

In the first order of $a$, unknown $\Sigma$ and $S$ are determined from the following commutators: 
\begin{equation}
[\Sigma , \gamma ]=\gamma I, \qquad i[Q,S]=\Sigma Q
\label{sigma_spin}
\end{equation} 
whose solutions are 
\begin{equation}
 \Sigma^{cd}=\frac{1}{4}\Big (\gamma^{c} \gamma^{d}-\gamma^{d} \gamma^{c}\Big),\qquad 
 S^{cd}=\frac{1}{2} \big (\,\overline{Q} \, \Sigma^{cd}\, Q\,\big )\label{sigma_s}  
 \end{equation} 
 where $\overline{Q}=Q^T \beta$. Because $\beta\, \Sigma^T=-\Sigma \, \beta$ one has 
 \begin{equation}
 \beta V^T=V^{-1}\beta
 \end{equation}
 which implies that  $Q^{\prime}=VQ$ also obeys the commutation relations  \eqref{commutators} and  $U^{-1} \,\overline{Q} \, U=\overline{Q}\, V^{-1}$. 
 
It is well known that  $\Sigma^{cd}$  obey Equation~\eqref{Lorentz_generators}, thus realising the  four-dimensional (spinor) representation of the Lie algebra of the Lorenz group.   As matrices $\beta \Sigma^{cd}$ are symmetric (cf.  (\ref{commutators})), it follows from \eqref{D} that spin operators $i S^{cd}$ from the above equation  obey the same commutator relations   \eqref{Lorentz_generators}, i.e., they also realise a representation of the Lie algebra of the Lorentz group. 

{The} explicit form of these operators is presented in \eqref{smn}.  These operators are a distinct feature of the new Dirac equation. When acting on  an $L_2$ function {$\Phi(x,q)$} depended on auxiliary variables $q_1$ and $q_2$ { (i.e., $\int_{-\infty}^{\infty} |\Phi(x,q)|^2 dq<\infty$ for arbitrary $x$)}, they lead to an  infinite-dimensional representation of the homogeneous Lorentz group.  A simple way to see it explicitly is to write the function as an infinite series in suitable functions of  $q_1$ and $q_2$. A convenient choice  is  eigenfunctions of  the harmonic oscillators $\hat{H}_1=\frac{1}{2}(q_3^2+q_1^2)$ and $\hat{H}_2=\frac{1}{2}(q_4^2+q_2^2)$ (the Hermite functions):  $\hat{H}_i\phi_m(q_i)=(m+\frac{1}{2} )\phi_m(q_i)$ (cf. \eqref{hermite})
\begin{equation}
\Phi(x,q)=\sum_{n,m=0}^{\infty} A_{mn}(x) \phi_m(q_1) \phi_n(q_2)\, .
\label{expansion_Hermite} 
\end{equation}
(Due to a symmetry of the new Dirac equation, integers $m$ and $n$ are of the same parity.)

Coefficients $A_{mn}(x)$ form an infinite number of wave function components of  Dirac particles. 
The action of operators $S^{cd}$ on  functions   $\phi_m(q_1)$ and  $\phi_n(q_2)$ induces a transformation of $A_{mn}$ corresponded to an infinite (and unitary) representation of the homogeneous Lorenz group (see, e.g.,  \cite{majorana, fradkin}, \cite{harish, gelfand}, and \cite{Dirac_III}).  Its explicit form can be obtained from \mbox{Equation~\eqref{functional_relations}.}  In a sense, an infinite ladder of coefficients $A_{mn}$ is hidden in the dependence of an eigenfunction on  variables $q_1$ and $q_2$, and function $\Phi(x,q)$ is a concise generating function of these coefficients.  

Spin matrices $\Sigma^{cd}$ equal  the commutator of two $\gamma$-matrices:  $\Sigma^{cd}=\tfrac{1}{4}[ \gamma^c, \gamma^d]$. Similarly, spin operators $S^{cd}$ also can be written as the commutator of two operators (cf. \eqref{D}, \eqref{comD}):
\begin{equation}
S^{cd}=i\big [\Gamma^c\,,\Gamma^d\big ],\qquad  \Gamma^c=-\frac{1}{4} \big ( \overline{Q}  \gamma^c Q \big ). 
\label{Gamma}
\end{equation}  

Operators $S^{cd}$ together with $\Gamma^c$ form a  representation of the group $SO(3,2)$ \cite{Dirac_IV, lectures} (see also Appendix~\ref{app_A}). 

From \eqref{covariance}, it follows that 
\begin{equation}
\Phi(x,q)=U(\Lambda)\, \Phi(\Lambda^{-1}x,q)\, .
\end{equation}

This invariance leads to the conservation of the total  momenta \cite{Dirac_I} 
$J^{cd}=M^{cd}-i S^{cd}$.  $M^{c d}=x^{c}\partial^{d}-x^{d}\partial^{c}$  are the usual momenta  and $S^{cd}$ represent the spin momenta. 


\section{Additional Equations}\label{additional_equations}

  {The} new Dirac equation \eqref{P_ops} is an overdetermined system of equations and any linear combination of these equations  is also a valid equation; thus, $M\hat{P}_j\, \Phi=0$ with any $4\times 4$ matrix $M$. Multiplying this equality by $\overline{Q} $ and summing over spin indices, one obtains $16$ equations quadratic {in}  $Q$: $\overline{Q} M (\gamma^{\mu} \partial_{\mu}-m)Q \Phi=0$.  When $M=\gamma^5$, one obtains identical zero and the remaining $15$ equations (corresponded to $M=1,\; \gamma^{\mu},\; \Sigma^{\mu \nu},\; \gamma^5 \gamma^{\mu} $, respectively) take the {following}  form \cite{Dirac_I,staunton}:
\begin{eqnarray}
&& \big (\Gamma^{\mu}\partial_{\mu}+\frac{i}{2} m  \big )\Phi=0\, ,\label{first}\\
&& \big (\frac{i}{2} \partial^{\mu} +S^{\mu \nu} \partial_{\nu}  +m\Gamma^{\mu}  \big )\Phi=0\, ,\label{second} \\
&&\big(  \Gamma^{\mu} \partial^{\nu}-\Gamma^{\nu} \partial^{\mu}+m S^{\mu \nu} \big )\Phi=0\, ,\\
&&\big ( e_{\mu \nu \alpha \beta} S^{\nu \alpha} \partial^{\beta} \big )\Phi=0\, .\label{last} 
\end{eqnarray} 

The first of these equations is {(up to a notation)} the positive-energy Majorana equation \eqref{old_majorana}  \cite{majorana},   and the last one coincides (up to a factor)  with the Pauli--Lubanski (pseudo)vector, indicating that field $\Phi$ has zero spin.  

The Majorana equation \eqref{first} is a characteristic feature of  positive-energy equations. All solutions of such equations, in a way or another, obey this equation. Though the form of the Majorana equation is similar to the usual Dirac equation \eqref{dirac}, they lead to different conclusions.  The main point is that, contrary to the usual $\gamma^0$ matrix,  which has positive and negative eigenvalues (see \eqref{usual_gamma}),  operator  $\Gamma^0$ has the form (cf. \eqref{Gamma_m}) 
\begin{equation}
\Gamma^0=\frac{1}{4}\Big ( (-\partial_{q_1}^2+q_1^2) +(-\partial_{q_2}^2+q_2^2) \Big).
\end{equation}

When acting on $L_2$ functions of $q_1, q_2$ {invariant over inversion, as in \eqref{expansion_Hermite}}, it has only positive eigenvalues equal to $\tfrac{1}{4} (2j+1)$ with integer $j\geq 0$. It is this property that implies that particles described by the Majorana equation have masses decreasing with the spin: $m(j)=m/(j+1/2)$.  

In Dirac's approach,   the above equations are just consequences of Equation~\eqref{P_ops}. It was argued in \cite{staunton} that Equation~\eqref{second} 
may be chosen as the fundamental one, 
 leading to the same plane wave solution as in \eqref{plane_wave}. A more general equation has been considered in that reference:
\begin{equation}
\Big (i \kappa\partial^{\mu} +S^{\mu \nu} \partial_{\nu}  +m\Gamma^{\mu}  \Big )\Phi=0
\label{kappa}
\end{equation} 
with a parameter $\kappa$. It has been shown that this system of equations is consistent only for two values  $\kappa=\tfrac{1}{2}$ and 
$\kappa=1$. 
The first one corresponds to the Dirac equation \eqref{second}, and the second one gives a new equation describing particles transforming by an infinite-dimensional representation of the Lorentz  group but with spin $\tfrac{1}{2}$ in the rest frame. Contrary to the Dirac particles described in this note, such spin-$\tfrac{1}{2}$ particles  permit the interaction  with electromagnetic fields. 


\section{No-Interaction Theorem}\label{no_interaction_theorem}

The principal goal of the new Dirac equation  was the construction of a relativistic equation with only positive energy solutions. But for the relation with dark matter {advocated} here, the most important is another unusual  fact: that Dirac particles cannot interact with electromagnetic fields.   

Usually, the interaction of a particle with an electromagnetic field $A_{\mu}$ is introduced within the so-called minimal coupling by the substitution $\partial_{\mu}\to D_{\mu}= \partial_{\mu} +i e A_{\mu}(x)$ into the wave equation.    The  remarkable propriety  of the new Dirac equation is that this standard procedure leads to the inconsistency of resulting equations, which means that that Dirac particles  cannot interact with electromagnetic fields (at least with minimal coupling). This fact  had been briefly mentioned in the very end of the Dirac paper \cite{Dirac_I} and has been discussed in detail in \cite{biedenharn} (see also \cite{mukunda}). For completeness, that reasoning is presented below. 

Let us consider the new Dirac equation with minimal coupling:
\begin{equation}
\big( \gamma^{\mu}D_{\mu}-m \big )Q\Phi (x,q) =0\, ,\qquad D_{\mu}= \partial_{\mu} +i e A_{\mu}(x)\, .
\end{equation}

Applying the operator $\big(\gamma^{\mu}D_{\mu}+m\big)$ to this equation, one finds 
\begin{equation}
\Big (\big (\eta^{\mu \nu} D_{\mu}D_{\nu}-m^2\big )+ie \Sigma^{\mu \nu} F_{\mu \nu} \Big)Q\Phi(x,q)=0
\label{no_interaction}
\end{equation}
where $F_{\mu \nu} =\partial_{\mu} A_{\mu} -\partial_{\nu} A_{\mu}$ and 
$\Sigma^{\mu \nu}=\frac{1}{4} (\gamma^{\mu}\gamma^{\nu} -\gamma^{\nu}\gamma^{\mu})$.

To  remove the first term in \eqref{no_interaction}, (which corresponds to the Klein--Gordon equation with the minimal coupling),  let us   
 multiply \eqref{no_interaction} by an antisymmetric matrix $G$ such that $\mathrm{Tr}(G\beta)=0$ and convolute the result with $Q^{T}$. According to \eqref{QGQ},  it will  cancel  the Klein--Gordon term. Besides $16$ independent products of the above $\gamma$ matrices, there are five matrices with such proprieties: $\beta \gamma^5 \gamma^{\mu}$ and $\beta \gamma^5$.  Let $G_j$ be one of these five matrices. After the indicated transformations, Equation~\eqref{no_interaction}  takes the form
\begin{equation}
F_{\mu \nu} h^{\mu \nu}_j \Phi=0\, ,\qquad h^{\mu \nu}_j= Q^{T} G_j \Sigma^{\mu \nu}Q
\end{equation}  
where $h^{\mu \nu}_j$ are certain operators constructed from ten possible symmetric bilinear combinations of $q_j$.  Using properties of $\gamma$-matrices (cf. \eqref{5_mu_nu} and \eqref{lambda_sigma}), one obtains 
\begin{equation}
\big (\overline{Q} \gamma^5 \gamma^{\lambda} \Sigma^{\mu \nu} Q\big )=\frac{1}{2}e^{\lambda \mu \nu \rho}\Gamma_{\rho}\, ,\qquad 
\big (\overline{Q} \gamma^5 \Sigma^{\mu \nu} Q\big )=\frac{1}{8} e^{\mu \nu \lambda \rho}S_{\lambda \rho} 
\end{equation}
where $S^{\mu \nu}$ are antisymmetric spin operators introduced in \eqref{sigma_s}), and $\Gamma^{\mu}$ is defined by \eqref{Gamma}. 

Therefore,  the  above equations take the form
\begin{equation}
\tilde{F}_{\mu \nu}\Gamma^{\nu} \Phi=0 \, ,\qquad \tilde{F}_{\mu \nu} S^{\mu \nu} \Phi=0\, .
\label{dual_F}
\end{equation}

Here, $\tilde{F}_{\mu \nu}=\frac{1}{2}e_{\mu \nu}^{\hspace{2ex} \rho \delta}F_{\rho \delta}$ is the dual electromagnetic tensor. 

In \cite{biedenharn}, it was argued  that from (\ref{dual_F}),  it follows that if $F_{\mu \nu} \neq 0$,  all operators $\Gamma^{\mu}$ and $S^{\mu \nu}=i\big [ \Gamma^{\mu},\, \Gamma^{\nu}\big ]$ annihilate $\Phi(x,q)$, which means that $\Phi(x,q)$ is independent on $q$, and consequently, the new Dirac equation with the minimal coupling of electromagnetic fields has no nontrivial solution. In other words, Dirac particles cannot interact with nonzero electromagnetic fields. 

Practically all works on this subject have stressed this 'drawback',  and many papers have been devoted to  generalisations of  the new Dirac equation, which  {would} permit the electromagnetic interaction. It appears that it can be carried out only by considerable complications of the equation, e.g., by considering  higher-order momentum terms \cite{staunton,biedenharn}, or by increasing the number of 'intrinsic' variables  \cite{chiang} or by introducing parabosonic constituents \cite{lectures,sudarshan}. 

{The} no-electromagnetic-interaction theorem has been considered in the past as a serious  disadvantage of the new Dirac equation,  and to a great exten{t},   it was responsible for the {loss}  of interest to this subject. But it is exactly what is required  for dark matter, which{, by definition,}   is a  substance whose characteristic property is that it does not interact with electromagnetic fields.   Dirac particles cannot be connected {to} electromagnetic fields not by simply postulating that they have zero charge but because their relativistic equation is inconsistent in the presence of the fields. It is this property that makes Dirac particles an ideal and  natural candidate for the illusive dark matter.  {In a sense, merely the existence of particles transforming  by an infinite-dimensional representation of the Lorenz group implies the existence of dark matter.}


\section{Gravity and Dirac Particles}\label{gravity_Dirac_particles}

 The  interaction of elementary particles with gravitational fields has attracted a certain attention from the very beginning of wave mechanics \cite{fock,schrodinger}.   Due to the smallness of the gravitational interaction, only very special  experiments are possible  at the present time (see, e.g., \cite{antigravity}), and such a subject is mostly of a pure theoretical interest.   But for Dirac particles, this problem is more fundamental, as they seem to interact directly only with the gravity  (as some other dark matter candidates). 

A usual way to describe spinors in gravitational fields (i.e., in a curved space)  is the tetrad formalism  (see, e.g., \cite{fock,weinberg}). In a nutshell,  it consists to  introduce  tetrads (vierbeins) $t_{\mu}^{a}(x)$ instead of (symmetric) metric tensor $g_{\mu \nu}(x)$,
 such that 
\begin{equation}
g_{\mu \nu}(x)=t_{\mu}^{\hspace{1ex} a}(x) t_{\nu}^{\hspace{1ex} b}(x)\eta_{a b} \, , \qquad \eta=\mathrm{diag}(-1,1,1,1)\, . 
\end{equation} 

For completeness, the main formulas {of the  tetrad formalism}  are presented in Appendix~\ref{app_tetrad}. Using the indicated rules, any given relativistic equation in a flat space (without gravity) can be transformed in a covariant manner  to  a curved space (with gravitational fields). 

However, the new Dirac equation is not  {just} one equation but an overdetermined system of equations, and it is not clear which of the many (almost) equivalent equations in a flat space (cf., \eqref{dirac}, \eqref{new_dirac} and \eqref{first}--\eqref{last}) have to be converted to a curved space.
 Different equations  may and will  lead to different answers. For example, it is claimed in \cite{ahner} that the use of   Equation~\eqref{kappa} with $\kappa=1$ (i.e., for spin-$\tfrac{1}{2}$ particles) leads to a consistent theory in gravitational fields. 

A natural way of introducing gravitational fields into  the new Dirac Equations \eqref{dirac} and \eqref{new_dirac} consists  of transforming these equations themselves to a curved space.  Using formulas from Appendix~\ref{app_tetrad}, it is easy to check that the transformed equations can be written in two  equivalent forms:
\begin{equation}
\big ( \hat{\gamma}^{\mu}Q D_{\mu}-m\,Q\big ) \Phi=0,\qquad \big (\hat{\gamma}^{\mu} D_{\mu}^{\prime} -m\big )Q \Phi=0
\end{equation}
where 
\begin{equation}
D_{\mu} =\partial_{\mu}+\frac{i}{2} \omega_{\mu ab} S^{ab}\, ,\qquad D_{\mu}^{\prime} =\partial_{\mu}+\frac{1}{2}\omega_{\mu ab} \big ( \Sigma^{ab}+ i  S^{ab}\big ) \, .
\end{equation}

As $\Sigma Q=i[Q ,S]$ (cf. \eqref{sigma_spin}), these two expressions are consistent. (For clarity, covariant derivatives acted on different quantities are denoted here by different {symbols}.)  

As above, one can derive many consequences of that equation. In particular, by acting on this  equation by $\big ( \hat{\gamma}^{\nu}D_{\nu}^{\prime}+m\big)$, one concludes that 
\begin{equation}
\Big (Q \big ( g^{\mu \nu}D_{\mu}D_{\nu}-m^2\big )  +\Sigma^{\nu \mu} Q \big [ D_{\nu},D_{\mu}\big ] \Big )\Phi=0\, . 
\end{equation}

Using \eqref{commutator_phi}, it can be rewritten as follows: 
\begin{equation}
\Big (Q \big ( g^{\mu \nu}D_{\mu}D_{\nu}-m^2\big )  +\frac{i}{2} \Sigma^{\mu \nu} \,Q \, R_{\nu \mu a b}S^{ab}  \Big )\Phi=0\, .
\label{dirac_gravity}
\end{equation}
where    $R_{\lambda \rho  a b} $ is the Riemann curvature tensor. 

This equation has a form similar to \eqref{no_interaction}, and the same arguments can be applied to cancel the first (scalar) term.   Therefore, the necessary conditions for the existence of solutions require that the following two expressions are zero (cf.  with Equation~\eqref{dual_F}):
\begin{equation}
e^{\mu \nu \lambda \rho} \Gamma^{\nu} R_{\lambda \rho  a b} S^{ab}\Phi=0\, ,\qquad e^{\mu \nu \lambda \rho}S^{\mu \nu} R_{\lambda \rho  a b} S^{ab}\Phi=0\, . 
\label{no_gravitation} 
\end{equation}

Contrary to the interaction with electromagnetic fields, these equations may have nontrivial solutions. The point is that $ R_{\lambda \rho  a b} S^{ab}\Phi$ is an antisymmetric tensor of the second degree. When the metric is without torsion, there is only one such tensor, $S_{\lambda \rho}$. It means that 
\begin{equation}
R_{\lambda \rho  a b} S^{ab}\Phi= S_{\lambda \rho}\,\hat{W} \Phi
\label{W}
\end{equation}
where $\hat{W}$ is a scalar operator. By convoluting the above equation with $S^{\lambda \rho}$ and using \eqref{identities_S}, one obtains

\begin{equation}
\hat{W} \Phi=-\frac{2}{3}S^{\lambda \rho}  R_{\lambda \rho  a b} S^{ab}\Phi\, . 
\end{equation}

 Substituting \eqref{W}  into \eqref{no_gravitation} and using identities \eqref{identities_epsilon} shows  that all 
 Equations~\eqref{no_gravitation} become automatically  zero. 
 
 Multiplying \eqref{dirac_gravity} by $\overline{Q}$, one finds that 
 \begin{equation}
 \Big ( g^{\mu \nu}D_{\mu}D_{\nu}-m^2 +\frac{1}{2} S^{\mu \nu}  R_{\nu \mu a b}S^{ab}  \Big ) \Phi=0\, . 
 \end{equation}
 
 It is plain that the same equation can be obtained directly by the transformation  Equation~\eqref{second} to the curved space and applying the covariant derivative to the result. 
 
 The corresponding equation for the usual Dirac equation has {been} derived   in \cite{schrodinger}, and it  has the similar form:
\begin{equation}
 \Big ( g^{\mu \nu}D_{\mu}D_{\nu}-m^2 +\frac{1}{2} \Sigma^{\mu \nu}  R_{\nu \mu a b}\Sigma^{ab}  \Big ) \Psi=0
 \end{equation} 
with the difference being that  $\Psi$ is a usual four component 
 spinor, and due to  \eqref{gamma_anti_com},  \eqref{symmetry_R}, and \eqref{bianchi},  the term with the Riemann tensor can be reduced to a scalar curvature $R=g^{\nu \lambda}g^{\mu \rho} R_{\nu \mu \lambda \rho}$
\begin{equation}
\frac{1}{2} \Sigma^{\mu \nu}  R_{\nu \mu a b}\Sigma^{ab}=-\frac{1}{4}R\, .
\end{equation} 

The above arguments demonstrate how  the main obstacle to the interaction of  Dirac {particles} with electromagnetic fields can be overcome  in gravitational fields, but unfortunately,   they do not prove the existence of its solutions. The author is not aware of  mathematical  results, which could be applied for such overdetermined equations, and further investigations are desirable.


\section{Conclusions}\label{conclusion}

 {The}  relativistic invariance implies  that elementary particles are transformed by irreducible representations of the Lorenz group. All known {elementary} particles correspond to \textit{finite}-dimensional representations. But \textit{infinite}-dimensional representations of the homogeneous  Lorenz group do exist, and it is natural to conjecture that  particles related with such representations exist as well. 
  
  The new Dirac equation describes the simplest of such  particles with zero spin in the rest frame. Originally, this equation has been  developed  as an example of  a relativistic wave equation whose all solutions 
   have only positive energies. Similar to all known positive-energy equations,  solutions of the new Dirac equation  obey also the infinite-dimensional Majorana equation. But the latter has too large a spectrum of possible solutions.  To select a single solution, the new Dirac equation is constructed  to be not an isolated  equation but an overdetermined system of equations such that one of its consequences is the Klein--Gordon equation with a fixed mass.  This  construction {necessarily}  requires that all these  different equations are consistent (i.e., have the same  solutions), which is a nontrivial statement. 
  
  A Dirac particle is described by  one function depended on  four usual space-time coordinates and two additional  variables. The later are only dummy variables  introduced in such a way that  the expansion over them generates an infinite number of particle wave components  corresponding to the infinite-dimensional representation of the Lorenz group. 
  
    It was proved  that the resulting equations have no common solution when electromagnetic fields are introduced into the new Dirac equation within the minimal coupling. It means that  Dirac particles cannot, in principle, interact with electromagnetic fields (with the minimal coupling) but it comes as a consequence of their internal structure and  has not been imposed \textit{ad hoc}. Consequently, if Dirac particles exist they  will {automatically} have dark matter properties. 

 The distinctive  property  of Dirac particles is that they have only positive energy which  infers the absence of Dirac antiparticles. Thus, Dirac particles   break the CPT theorem but without violating any fundamental law. It {signifies} that  such particles might be  a new source of observed matter--antimatter asymmetry. 
 
Dirac particles (as well as  other  particles transformed by infinite-dimensional representations of the homogeneous Lorentz group) are unusual, interesting, and potentially important {objects}, which  may be  a  clue  to a new physics. Their characteristic  properties  are peculiar,  and to a large extent,  counterintuitive. To summarise: 
\begin{itemize}
\item They have spin zero (in the rest frame) but are not scalars.
\item They obey covariant relativistic  equations but have only positive energies, and {consequently} do not have antiparticles.
\item Their equations are linear in the momenta but do not permit the minimal coupling to electromagnetic fields.   
\end{itemize}

Unfortunately, Dirac particle  investigations  are undeveloped. No usual textbooks have mentioned their existence,  no standard models incorporate  them,  and  many important questions remain unanswered. At the present stage of the study, it seems that Dirac particles are sterile and can interact directly only with the gravity.   Detailed examinations  of the interaction of Dirac particles with usual matter  through gravitational loops and  their  production during the Big Bang are essential  to give them the status of established elementary particles capable of explaining dark matter. 

Such inquiries   are not straightforward and require the development of new methods,  but the possibility that Dirac particles could dominate the universe clearly stresses  the importance of the topic  and the necessity to investigate it seriously. 
\vspace{6pt}

\acknowledgments{The author is grateful to M. Cirelli for useful comments.}

\appendix
\section{Useful Relations}\label{app_A}
\begin{itemize}

\item  The choice in \cite{Dirac_I} is equivalent  to the following gamma matrices obeying (\ref{condition}): $\gamma^0=\beta$,     
\begin{equation}
\gamma^1=\Big ( \begin{array}{cc} {-\sigma_3}& {0}\\ {0}& {\sigma_3} \end{array}\Big)\, , \quad  
 \gamma^2=\Big ( \begin{array}{cc} {\sigma_1}& {0}\\ {0}& {-\sigma_1} \end{array}\Big)\, ,\quad 
\gamma^3=\Big ( \begin{array}{cc} {0}& {-1}\\ {-1}& {0} \end{array}\Big )\, ,
\label{usual_gamma} 
\end{equation}
and
\begin{equation}
\gamma^5=\gamma^0 \gamma^1\gamma^2\gamma^3=\Big ( \begin{array}{cc} {i\sigma_2}& {0}\\ {0}& {-i\sigma_2} \end{array}\Big)
\end{equation}
where $\sigma_i$ are the {usual} $2\times 2$ Pauli matrices
\begin{equation}
\sigma_1=\Big ( \begin{array}{cc} {0}& {1}\\ {1}& {0} \end{array}\Big)\,, \quad  
 \sigma_2=\Big ( \begin{array}{cc} {0}& {-i}\\ {i}& {0} \end{array}\Big)\, ,\quad 
\sigma_3=\Big ( \begin{array}{cc} {1}& {0}\\ {0}& {-1} \end{array}\Big )\,.
\end{equation} 

\item When working with quadratic operators  {in} $Q$, the following two formulas are useful. 
\begin{itemize}
\item[(a)]  For  an antisymmetric matrix $G_{jk}=-G_{kj}$ there is an operator identity
\begin{equation}
Q^{T} G Q\equiv q_{j} G_{jk}q_k=-\frac{i}{2}\mathrm{Tr}(G\beta)\, .
\label{QGQ}
\end{equation}
\item[(b)]  Let  us consider the following transformation of a $4\times 4$ matrix $A$ (see, e.g., \cite{lectures}):
\begin{equation}
\mathcal{D}: A\to \mathcal{D}(A)=\frac{i}{2} \overline{Q}  A Q \, , \qquad \overline{Q}=Q^T\beta\, .
\label{D}
\end{equation}
If   $\beta A$ and $\beta B$ are  symmetric matrices,  then the direct calculations prove that this transformation  preserves commutator relations
\begin{equation}
[\mathcal{D}(A),\mathcal{D}(B)]=\mathcal{D}([A,B])\ .
\label{comD}
\end{equation}
\end{itemize}
\item The spin operators, which realise the infinite-dimensional representation of the Lorentz group, are  defined in \eqref{sigma_s}: $S^{cd}=\frac{1}{2} \overline{Q} \Sigma^{cd}Q$. Rewriting them by components gives
\begin{eqnarray}
S^{01}&=&\frac{1}{4} (q_1^2-q_3^2-q_2^2+q_4^2)\, ,\qquad S^{02}=\frac{1}{2}(q_3q_4-q_1 q_2)\, ,\nonumber\\ 
S^{03}&=& \frac{1}{2}(q_1 q_3+q_4 q_2)\, ,\qquad 
S^{12}=\frac{1}{2}(q_2 q_3-q_1 q_4)\, ,\\
S^{13}&=&\frac{1}{4}(q_2^2+q_4^2-q_1^2-q_3^2),\qquad S^{23}=\frac{1}{2}(q_1q_2+q_3q_4)\, .\nonumber
\label{smn}
\end{eqnarray}
\item The {components of} vector operator $\Gamma^c=-\frac{1}{4} \overline{Q}  \gamma^c Q$ {are} 
\begin{eqnarray}
\Gamma^0&=&\frac{1}{4}(q_1^2+q_2^2+q_3^2+q_4^2)\, ,\quad \Gamma^1=\frac{1}{2}(q_2q_4-q_1q_3)\, , \label{Gamma_m}\\
\Gamma^2&=&\frac{1}{2} (q_1q_4+q_2q_3)\, ,
\qquad \Gamma^3=\frac{1}{4}(q_1^2+q_2^2-q_3^2-q_4^2)\, .\nonumber
\end{eqnarray}
\item In the calculations, the following identities are also  helpful (see, e.g.,  \cite{lectures}): 
\begin{equation}
\gamma^5 \Big (\gamma^{\mu}\gamma^{\nu}-\gamma^{\nu}\gamma^{\mu} \Big )=e^{\mu \nu}_{\hspace{2ex} \rho \sigma}\gamma^{\rho}\gamma^{\sigma}
\label{5_mu_nu}
\end{equation}
and 
\begin{equation}
\frac{1}{2}\gamma^{\lambda} \Big (\gamma^{\mu}\gamma^{\nu}-\gamma^{\nu}\gamma^{\mu} \Big )=
\eta^{\lambda \mu}\gamma^{\nu}-\eta^{\lambda \nu} \gamma^{\mu}+e^{\lambda \mu \nu}_{\hspace{3ex}  \rho} \gamma^5 \gamma^{\rho}\, .
\label{lambda_sigma}
\end{equation}
Here, $e^{\mu \nu \rho \sigma}$ is {the completely} antisymmetric tensor with $e^{0 1 2 3}=1$.

\item For convenience, the commutators of spin operators $S$ and $\Gamma$ are listed below:
\begin{eqnarray}
 \big [ S^{cd},\, S^{kr} \big ]&=& i \big ( \eta^{c k} S^{d r} + \eta^{d r} S^{c k}  -\eta^{d k} S^{c r} -\eta^{c r} S^{d k} \big )\, ,\\
\big[ \Gamma^c,\, \Gamma^d\big ]&=&-iS^{cd}\, ,\\
\big [  S^{cd},\, \Gamma^b \big ]&=& i \big ( \eta^{c b} \Gamma^d-\eta^{d b} \Gamma^c \big )\, . 
\end{eqnarray}
These commutators are the periphrases of {the} corresponding relations for the usual $\gamma$-matrices.  

\item The above operators obey a number of  identities (see e.g., \cite{staunton}):
\begin{eqnarray}
&&S_{ab}\,S^{ab}=-\frac{3}{2},\qquad \Gamma_{a}\, \Gamma^a=\frac{1}{2}\, ,\label{identities_S}\\
&&e^{abcd} S_{ab}\,S_{cd}=0,\qquad e^{abcd} S_{ab}\,\Gamma_{c}=0\, ,\label{identities_epsilon}\\
&&\Gamma_a \,S^{ab}=-\frac{3i}{2} \Gamma^b,\qquad  S^{ab} \,\Gamma_b=-\frac{3i}{2} \Gamma^a,\\ 
&&S^a_{\hspace{1ex} b}\,S^{bc}=\frac{1}{2} \big ( \eta^{ac}-2\Gamma^a\, \Gamma^c-3iS^{ac} \big )\, .
\end{eqnarray}

\item Let us introduce a $5\times 5$ metric tensor $\eta^{A B}=\mathrm{diag}(-1,1,1,1-1)$ with $A,B=0,\ldots 4$ and denote $S^{4\,a}=\Gamma^a$. {Ten} antisymmetric operators  $S^{ab}$  and $S^{4\,a}$ with  $a,b=0,\ldots 3$ form a representation of the group O($3,2$) as {the} above commutation relations can be rewritten in the form {(cf. with Equation~\eqref{Lorentz_generators})}
\begin{equation}
 \big [ S^{ CD},\, S^{KR} \big ]= i \big ( \eta^{C K} S^{D R}+ \eta^{D R} S^{C K}  -\eta^{D K} S^{C R} -\eta^{C R} S^{D K} \big )
\end{equation}
where majuscule letters are from $0$ to $4$.

\item The Hermite functions used in  the expansion \eqref{expansion_Hermite} have the form (see, e.g., \cite{bateman})
\begin{equation}
\phi_n(q)=\Big ( 2^n n! \sqrt{\pi} \Big )^{-1/2} H_n(q)e^{-q^2/2}
\label{hermite}
\end{equation}
where $H_n(q)$ are {the}  Hermite polynomials. These functions are orthogonal: 
\begin{equation}
\int_{-\infty}^{\infty} \phi_m(q) \phi_n(q)dq=\delta_{mn}
\end{equation}
and obey the following relations:
\begin{eqnarray}
q \phi_n(q)&=&\sqrt{\frac{n}{2}} \phi_{n-1}(q)+\sqrt{\frac{n+1}{2}} \phi_{n+1}(q)\, ,\label{functional_relations}\\
\phi_n^{\prime}(q)&=&\sqrt{\frac{n}{2}} \phi_{n-1}(q)-\sqrt{\frac{n+1}{2}} \phi_{n+1}(q)\, . \nonumber
\end{eqnarray}
{The use of these identities  permits one to rewrite any equation in auxiliary variables $q_j$ in the form of infinite-dimensional matrices  without these variables.} 
\end{itemize}

\section{Tetrad Formalism }\label{app_tetrad}

A tetrad (vierbein) is, by definition, a set of  (usually orthogonal) vectors $t^{\mu}_{a}(x)$ defined in all space points $x$ which transform the true symmetric metric tensor  of a curved space $g^{\mu \nu}(x)$ locally into the flat  Minkowski metric tensor  $\eta=\mathrm{diag}(-1,1,1,1)$ (see, e.g., \cite{weinberg})
\begin{equation}
\eta_{a b}=t^{\mu}_{a}(x) t^{\nu}_{b}(x)g_{\mu \nu}(x)\, . 
\end{equation}

 Here Latin, (resp., Greek) letters indicate flat (resp., curved) coordinates.

The curved-space gamma matrices are determined as follows:
\begin{equation}
\tilde{\gamma}^{\mu}(x)=t^{\mu}_a(x) \gamma^a\, . 
\label{curved_gamma}
\end{equation} 

These $4\times 4$ matrices  obey the relation
\begin{equation}
\tilde{\gamma}^{\mu}(x) \tilde{\gamma}^{\nu}(x)+\tilde{\gamma}^{\nu}(x) \tilde {\gamma}^{\mu}(x)=2g^{\mu \nu}(x)\, .
\label{gamma_anti_com}
\end{equation}

For clarity,  all quantities multiplied by tetrads will be indicated by {the same letter as in the flat space but with}  a tilde: 
\begin{equation}
\tilde{\Gamma}_{\mu}(x)=t^{\mu}_a(x) \Gamma^{a}\, ,\qquad \tilde{\Sigma}^{\mu \nu}(x)=t^{\mu}_a(x) t^{\nu}_b(x) \Sigma^{ab}\, , \qquad \tilde{S}^{\mu \nu}(x)=t^{\mu}_a(x) t^{\nu}_b(x) S^{ab}\, .
\end{equation}

At a fixed point,  a (symmetric) metric tensor is determined by ten components, but there exist sixteen tetrads. 
 The remaining  six free parameters correspond to six-parameter local Lorentz transformations, which preserve  the Minkowski metric tensor. 

It is well known that to conserve a local invariance, it is necessary to change {the} usual derivatives to  covariant ones:
\begin{equation}
\partial_{\mu}\Psi(x)\longrightarrow D_{\mu}\Psi(x)=\big (\partial_{\mu}+C_{\mu})\Psi(x)
\end{equation}   
where $C_{\mu}$ is {a} connection whose explicit form depends on a quantity $\Psi(x)$ on which the derivative acts.   In general, if under the infinitesimal local Lorentz transformations  $x^{\mu}= x^{\mu}+a^{\mu}_{\nu}(x) x^{\nu} $ (cf. \eqref{Lorentz_transform}) {a field  $\Psi(x)$ transforms in the following way }  
$\Psi(x)\to \Psi(x)+a_{ab}(x)\hat{s}^{ab} \Psi (x)$, then  
\begin{equation}
C_{\mu}=\omega_{\mu ab}\,\hat{s}^{ab}
\end{equation}
with certain  coefficients $\omega_{\mu ab}$ depending on quantities considered. It is common to denote a covariant derivative by $D_{\mu}$, but a care should be taken when its explicit expression is used in calculations.  For vector (and tensor)  fields,  $\omega_{\mu ab}=\Gamma_{\mu ab}$, and   it is called  the Christoffel symbol.  Different cases  are indicated in Table~\ref{table_I}. 

\begin{table}
\caption{Covariant  derivatives of different quantities.\label{table_I}}
\begin{tabular}{|c|c|c|}
\hline
Contravariant Lorentz vector & $V^{\mu}\to V^{\mu}+a^{\mu}_{\lambda}(x)V^{\lambda}$ & 
$D_{\nu}V^{\mu}=\partial_{\nu}V^{\mu} +\Gamma^{\mu}_{\nu \lambda}V^{\lambda}$ \\ 
\hline
Contravariant tetrad vector  & $V^{b}\to  V^b+a^{b}_{c} (x) V^c $ & 
$D_{\nu} V^b=\partial_{\nu} V^b+\omega_{\nu c}^{\hspace{2ex} b} V^c $\\ 
\hline
Usual Lorentz  spinor & $\Psi\to \Psi +\frac{1}{2} a_{ab}(x) \Sigma^{ab} \Psi$ & $D_{\nu} \Psi=\big (\partial_{\nu} +\frac{1}{2} \omega_{\nu ab} \Sigma^{ab} \big )\Psi   $\\  
\hline
Dirac particle & $\Phi\to \Phi+\frac{i}{2} a_{ab}(x) S^{ab} \Phi $ & $D_{\nu} \Phi=\big( \partial_{\nu}+\frac{i}{2} \omega_{\nu ab } S^{ab} \big )\Phi $
\\ 
\hline
\end{tabular}

\end{table}

For covariant quantities, the connections  change the sign. When tensors are considered, it is necessary to add the corresponding connections for each index. For matrices (and operators) {such transformation}  corresponds to the commutator with the connections. 

Covariant derivatives  transform in the same way as the corresponding quantities in the flat space. In particular, they have to commute with the rising and lowering indices with the metric tensor. Therefore, 
\begin{equation}
D_{\nu} g_{\mu \lambda}=0\, , \qquad \partial_{\nu} g_{\mu \lambda}-\Gamma_{\nu \mu}^{\beta} g_{\beta \lambda} -\Gamma_{\nu \lambda}^{\beta} g_{\mu \beta}=0\, . 
\end{equation} 

This relation implies that   
\begin{equation}
\Gamma^{\mu}_{\nu \lambda}=\frac{1}{2} g^{\mu \rho} \left (  \partial_{\nu} g_{\rho \lambda} +\partial_{\lambda} g_{\rho \nu} -
 \partial_{\rho} g_{\mu \nu} \right ) ,\qquad \Gamma^{\mu}_{\nu \lambda}=\Gamma^{\mu}_{ \lambda \nu}.
 \end{equation}
 
Similarly, it is convenient to require that  
\begin{equation}
D_{\nu} t_{\mu}^{\hspace{1ex} a}=0\, ,\qquad \partial_{\nu} t_{\mu}^{\hspace{1ex} a}+\Gamma_{\nu \mu}^{\lambda} +\omega_{\nu c}t_{\mu}^{\hspace{1ex} c}=0 
\end{equation}
which leads to 
\begin{equation}
\omega_{\nu  c}^{\hspace{2ex} b}=t^{\mu}_c \big (- \partial_{\nu}t_{\mu}^b +\Gamma^{\lambda}_{\nu \mu} t^b_{\lambda} \big )\, , \qquad 
\omega_{\mu c b}=-\omega_{\mu b c }\, .
\label{xi}
\end{equation}

The same answer can be obtained from 
\begin{equation}
D_{\nu}\hat{\gamma}^{\mu}(x)=0\, ,\qquad  \partial_{\nu}\hat{\gamma}^{\lambda}(x)  +\Gamma_{\nu \lambda}^{\mu}\hat{\gamma}^{\lambda}(x)+\big [ g_{\nu}, \hat{\gamma}^{\mu}(x)\big ]=0 \, ,  \qquad g_{\mu}= \frac{1}{2} \omega_{\mu ab} \Sigma^{ab}\, . 
\end{equation}

In the same way, 
\begin{equation}
D_{\nu}\hat{\Gamma}^{\mu}(x)=0\, ,\qquad  \partial_{\nu} \hat{\Gamma}^{\lambda}(x) +\Gamma_{\nu \lambda}^{\mu} \hat{\Gamma}^{\lambda}(x)+\big [ G_{\nu}, \hat{\Gamma}^{\mu}(x)\big ]=0 \, ,\qquad  
G_{\mu}=\frac{i}{2} \omega_{\mu ab } S^{ab}\, .
\end{equation}

By definition, the commutator of two covariant derivatives acting at a covariant  vector is determined by  the Riemann curvature tensor $R_{ \nu \mu \rho  }^{\hspace{3ex}\lambda}$
\begin{equation}
\big [D_{\mu} ,\, D_{\nu}\big ] A_{\rho}=R_{ \nu \mu \rho  }^{\hspace{3ex}\lambda} A_{\lambda}\, ,\qquad   
R_{ \nu \mu \rho  }^{\hspace{3ex}\lambda}= \partial_{\nu} \Gamma_{\mu \rho}^{\lambda}  -  \partial_{\mu} \Gamma_{\nu  \rho}^{\lambda}
+ \Gamma_{\mu \rho}^{\sigma}  \Gamma_{\nu  \sigma}^{\lambda}- \Gamma_{\nu  \rho}^{\sigma}  \Gamma_{\mu \sigma}^{\lambda} \, . 
\end{equation}

It is known that the Riemann tensor obeys the symmetry relations (see, e.g., \cite{weinberg})
\begin{equation}
R_{\nu \mu \lambda \rho}=-R_{\mu \nu \lambda \rho}=-R_{\nu \mu \rho \lambda}=R_{\lambda \rho \nu \mu } 
\label{symmetry_R}
\end{equation}
and the Bianchi identity
\begin{equation}
R_{\nu \mu \lambda \rho}+R_{\nu \rho \mu \lambda }+R_{\nu \lambda \rho \mu }=0\, .
\label{bianchi}
\end{equation}

 As $D_{\nu} t_{\rho}^{\hspace{1ex} a}=0$  it follows  that  $\big [D_{\mu} ,\, D_{\nu}\big ]t_{\rho}^{\hspace{1ex} a}=0$,
from which one proves   the validity of the following identity: 
\begin{equation}
\partial_{\mu} \omega_{\nu b}^a- \partial_{\nu} \omega_{\mu b}^a+\omega_{\mu c}^{a} \omega_{\nu b}^c - \omega_{\nu c}^{a}
\omega_{\mu b}^c = -R_{\nu \mu  \rho  }^{\hspace{3ex} \lambda}\, t_{\lambda}^{\hspace{1ex} a}\, t^{\rho}_{\hspace{1ex} b}\,  .
\label{commutator_two_D} 
\end{equation}

Using this expression, one {obtains}
\begin{equation}
\big [ D_{\mu}, \, D_{\nu}\big ]\Phi=\frac{i}{2} R_{\nu \mu ab}S^{ab} \Phi\, .
\label{commutator_phi}
\end{equation}


\end{document}